\documentclass[reprint,superscriptaddress,amsmath,amssymb,aps,pra,floatfix]{revtex4-1}

\usepackage{amsfonts}
\usepackage{amsmath}
\usepackage{amsthm}
\usepackage{mathrsfs}
\usepackage{amscd}
\usepackage{amssymb}
\usepackage{subfigure}
\usepackage{amsxtra}
\usepackage{dcolumn}% Align table columns on decimal point
\usepackage{bm}           % bold math
\usepackage{bbm}
\usepackage{graphicx}
\usepackage{epstopdf}

\usepackage[colorlinks]{hyperref}

\newcommand{\ud}{\,\mathrm{d}}

\DeclareMathOperator{\tr}{tr}

\newcommand{\cF}{\mathcal{F}}
\newcommand{\cU}{\mathcal{U}}
\newcommand{\cE}{\mathcal{E}}

\newcommand{\bra}[1]{\ensuremath{\langle#1|}}
\newcommand{\ket}[1]{\ensuremath{\left|#1\right\rangle}}

\newcommand{\ketbra}[2]{\ensuremath{\left| #1 \right\rangle \left\langle #2 \right|}}

\newcommand{\TFI}{\mathcal{F}^{\mathrm{th}}}
\newcommand{\QFI}[1]{\mathcal{F}_{#1}}
\newcommand{\diff}[1]{\frac{\partial #1}{\partial \bar{n}}}
\newcommand{\brac}[1]{\left(#1\right)}

\begin{document}
\title{Surpassing the Thermal Cram\'er-Rao Bound with Collisional Thermometry}

\author{Angeline Shu}
\affiliation{Department of Physics, National University of Singapore, 2 Science Drive 3, Singapore 117542, Singapore}

\author{Stella Seah}
\affiliation{Centre for Quantum Technologies, National University of Singapore, 3 Science Drive 2, Singapore 117543, Singapore}
\affiliation{Department of Physics, National University of Singapore, 2 Science Drive 3, Singapore 117542, Singapore}

\author{Valerio Scarani}
\affiliation{Centre for Quantum Technologies, National University of Singapore, 3 Science Drive 2, Singapore 117543, Singapore}
\affiliation{Department of Physics, National University of Singapore, 2 Science Drive 3, Singapore 117542, Singapore}

\date{\today}

\begin{abstract}
In collisional thermometry, a system in contact with the thermal bath is probed by a stream of ancillas. Coherences and collective measurements were shown to improve the Fisher information in some parameter regimes, for a stream of independent and identically prepared (i.i.d.) ancillas in some specific states [Seah et al., Phys. Rev. Lett. \textbf{123}, 180602 (2019)]. Here we refine the analysis of this metrological advantage by optimising over the possible input ancilla states, also for block-i.i.d.~states of block size $b=2$. For both an indirect measurement interaction and a coherent energy exchange channel, we show when the thermal Cram\'er-Rao bound can be beaten, and when a collective measurement of $N>1$ ancilla may return advantages over single-copy measurements. 
\end{abstract}
\maketitle{}

\section{Introduction}
Quantum metrology aims to exploit quantum resources such as entanglement\cite{ Giovannetti2006,Maccone2013,Huang2016} and coherence\cite{Micadei2015,Pires2017,Castellini2019} to enhance measurement protocols beyond the classical limit. One of the more recent focus is the application of the framework of parameter estimation in the context of thermometry\cite{Jevtic2015,Correa2016,Hovhannisyan2018,Pasquale2018,Mehboudi2019,Seah2020,Bouton2020,Gebbia2020}.

In classical thermometry, a thermometer thermalized to its environment would have a precision described by the thermal Cram\'er-Rao bound\cite{Pasquale2018,Mehboudi2019}, which is dependent on its heat capacity. Once we move to the quantum regime, the use of miniaturized systems as measurement probes becomes desirable due to its reduced perturbative effects on the environment. However, this also generally means a lower precision due to the smaller heat capacity. The hope is therefore to look for quantum advantages that could arise from collective measurements of multiple probes such that we outperform the thermal Cram\'er-Rao bound.

In this paper, we consider a collisional thermometry model where a system acts as an intermediary of the environment through which a stream of i.i.d probes (or ancillas) measure the temperature. This was first proposed in Ref.~\cite{Seah2020} where it was shown that it was possible to surpass the thermal Cram\'er-Rao bound from measurement of a single ancilla. Here, we explore the optimal strategies to beat this bound by considering in Sec.~\ref{sec:ZZ} a standard indirect measurement interaction and in Sec.~\ref{sec:exchangeint} a coherent energy exchange channel between the system and probes. We show that it is indeed possible to significantly surpass the thermal Cram\'er-Rao bound with both interactions, especially at high temperatures. The advantage of the indirect measurement interaction lies primarily in the fact that the optimal strategy is straightforward and independent of the temperature measured. However, the enhancements possible via the energy exchange channel surpass that of the indirect measurement even if the most optimal strategy requires knowledge of the rough range of temperature of the environment. In both cases, there is observable advantage with collective measurements. Nevertheless, this effect becomes less significant in the limit of large number of ancillas.

\begin{figure}
\centering
\includegraphics[width=\columnwidth]{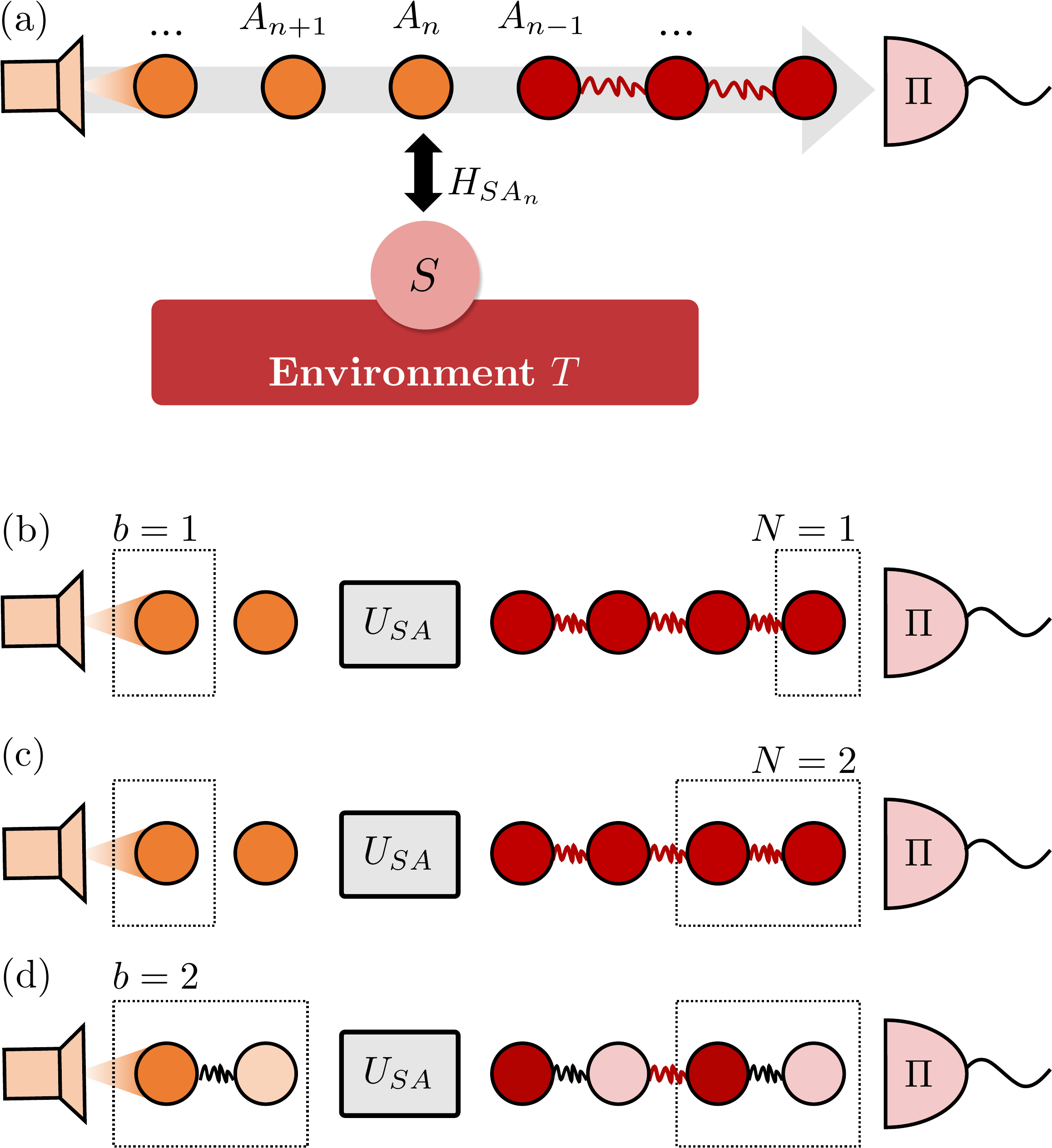}
\caption{\label{fig:sketch}
(a) Schematic diagram of the collisional thermometry protocol where a stream of ancillas interacts sequentially with a system $S$ in contact with a thermal reservoir at temperature $T$. (b)-(d) The system $S$ is probed by incoming ancilla states (orange) prepared in block sizes $b=1,2$ and the outgoing ancillas (red) are measured in block sizes $N=1,2$ after the interaction. The interaction $U_{SA}$ can build up correlation (red lines) even if the initial ancillas are uncorrelated.}
\end{figure} 

\section{Framework}
\subsection{Collisional thermometry model}

We consider a collisional thermometry protocol \cite{Seah2020} as sketched in Fig.~\ref{fig:sketch}. A system $S$ in contact with a thermal reservoir (``environment'') at temperature $T$ interacts with a stream of ancillas $\{A_1,A_2,\dots\}$ one at a time. One then measures the ancillas to gain information about $T$. There is a lot of freedom in such a model. For definiteness, in this paper we work under the following assumptions:
\begin{enumerate}
    \item The system and the ancillas are qubits with bare Hamiltonians $H_S$ and $H_{A_n}$.
    \item The stream of ancillas is periodic, each ancilla arriving every time $\tau$. Collisional models with Poisson-distributed arrival times have been considered, though not in the context of thermometry\cite{Pellegrini2009,Strasberg2018,Seah2019}
    \item The ancillas are prepared in block-i.i.d.~pure states $\Psi_{A_1...A_b}=\ket{\psi}\bra{\psi}$, with block size $b$ (for the results, we shall focus on $b=1$ and $2$).
    \item The coupling of the system with each ancilla is described by the same two-body Hamiltonian $H_{SA_n}$.
    \item The interaction of the system with each ancilla lasts a definite time $\tau_{SA}$, while in the remaining time $\tau-\tau_{SA}\equiv \tau_{SE}$ the system is left only in contact with the thermal environment. The interaction with the ancilla is deemed strong enough, such that the effect of the environment can be neglected during $\tau_{SA}$.
    
\end{enumerate}
Thus, the dynamics of the interaction of the system with the $(n+1)$-th block of ancillas reads 
\begin{equation}
    \rho_{SA_1\ldots A_b}((n+1)b) = \mathcal{C} \left[\rho_S(nb) \otimes \Psi_{A_1...A_b}\right] \label{eq:rhoSA}
\end{equation} with
\begin{equation}
    \mathcal{C}= \mathcal{E}_S \circ \; \mathcal{U}_{SA_b} \circ \ldots \circ \; \mathcal{E}_S \circ \; \mathcal{U}_{SA_1}\,,
\end{equation} where $\cU_{SA_{n}}[\rho] = U_{SA_n} \rho U_{SA_n}$ denotes the unitary evolution induced by the total Hamiltonian $H = H_S + H_{A_n}+ H_{SA_n}$, where $H_{S,A_n}$ are the bare system and ancilla Hamiltonians and $\cE_S[\rho]$ is the thermal map describing the coupling of the system and the environment. 

In the limit of weak system-environment coupling, the $SE$ interaction can be described by a Born-Markov master equation\cite{Lindblad1975,Strasberg2018,Seah2019}
\begin{align}
	\frac{\ud\tilde\rho}{\ud t}= \gamma(\bar{n}+1)\mathcal{D}[\sigma^-_S]\tilde\rho+\gamma\bar{n}\mathcal{D}[\sigma^+_S]\tilde\rho:=\mathcal{L}[\tilde\rho]\,,\label{eq:ME}
\end{align}
in the interaction picture where $\tilde\rho = e^{i H_S t}\rho e^{-i H_S t}$ and $H_S\propto\sigma^z$. Here, $\mathcal{D}[L]\rho=L\rho L^{\dagger}-\frac{1}{2}\{L^{\dagger}L,\rho\}$, $\gamma$ is the temperature-independent coupling constant and $\bar{n} = 1/[\exp(\hbar\Omega/k_{\rm B}T)-1]$ is the mean bath occupation number at frequency $\Omega$ and temperature $T$. The thermal map is then given by $\cE_S = e^{\mathcal{L} \tau_{SE}}$ and would describe relaxation of the system to a Gibbs state $\rho_S^{\rm th} \propto e^{-H_S/k_{\rm B}T}$.

For each block of ancillas, the reduced state $\rho_S$ of the system evolves stroboscopically according to 
\begin{equation}\label{eq:rhoSn}
    \rho_S((n+1)b) = \tr_{A_1...A_b}\left\{\mathcal{C} \left[\rho_S(nb) \otimes \Psi_{A_1...A_b}\right]\right\} =: \Phi\big[ \rho_S(nb)\big]\,.
\end{equation}
We shall focus on the steady state operation where $\rho_S^* = \Phi\big[ \rho_S^*\big]$. In this case, the state of the ancillas after the interaction is also the same for each block, given by $\rho^*_{A_1...A_b} = \tr_{S}\left\{\mathcal{C} \left[\rho_S^* \otimes \Psi_{A_1...A_b}\right]\right\}$.

\subsection{Temperature estimation}

The amount of information one can extract about a parameter $T$ from a state $\rho$ is described by first doing a positive operator valued measurement (POVM) $\Pi_{x}$ on the the state $\rho$. $x$ here labels the outcome of the measurement. From the resulting probability distribution $p(x)=\mathrm{tr}\{\Pi_{x}\rho\}$ one can compute the classical Fisher information (CFI)
\begin{align}
	F(\Pi,T,\rho)=\sum_{x}p(x)\left(\frac{\partial }{\partial T}\ln p(x)\right)^{2}.
\end{align}
The quantum Fisher information (QFI) is then obtained by optimizing over all possible POVMs, thereby returning
\begin{align}
	\QFI{}(T,\rho)=\max_{\Pi}F(\Pi,T,\rho)=\mathrm{tr}(\rho\Lambda^{2})\,,
\end{align}
where $\Lambda$ is the symmetric logarithmic derivative, obtained from solving the Lyapunov equation $\Lambda\rho+\rho\Lambda=2\partial_{T}\rho$. The reciprocal of the Fisher information sets a lower bound on temperature variance (\textit{Cram\'er-Rao bound})
\begin{align}\label{eq:cramerrao}
	(\Delta T)^{2}\geq \frac{1}{\QFI{}(T,\rho)}\,
\end{align}
valid regardless of the choice of estimator used to compute $T$.

Specifically for measurement of temperature, a canonical benchmark is the \textit{thermal Fisher information} $\TFI \equiv\cF(T,\rho_S^{\rm th})$ which is the quantum Fisher information computed for a thermalised probe. One can show that the thermal Fisher information for a qubit with frequency gap $\Omega$ is 
\begin{equation}
    \TFI = \frac{1}{\bar{n}(\bar{n}+1)(2\bar{n}+1)^2} \left(\frac{\partial \bar{n}}{\partial T}\right)^2\label{eq:TFI}
\end{equation}
where $\bar{n}$ is the mean bath occupation number at frequency $\Omega$. The corresponding lower bound is called \textit{thermal Cram\'er-Rao bound}.

In our work, the state $\rho$ that is to be measured for temperature estimation is the state $\rho^*_{A_1...A_b}$ of the ancillas after the interaction with the system, in the steady state regime for the system. In what follows, we omit the dummy bracket $(T,\rho)$ from the Fisher information, and rather keep track of the \textit{initial state of the ancillas} in the superscript and of \textit{the number of ancillas that are collectively measured} in the subscript. For instance, the expression $\QFI{N}^{\ket{g}}$ indicates that the ancillas were prepared in the single-particle ground state $\ket{g}$ (i.e.~$b=1$), and the information was extracted by a measurement on $N$ successive ancillas after the interaction (thus notice that $N$ is not equal to $b$ in general), see Fig.~\ref{fig:sketch}. For the special case of the thermal state, since the bare Hamiltonian of the ancilla is single-particle and thus the thermal state of $N$ ancillas is product, we have $\TFI_N=N\TFI_1\equiv N\TFI$. Here, we also define
\begin{equation}
    \cF_{N,b}^{\rm opt} = \max_{\psi} \cF_N^{\ket{\psi}}
\end{equation}
where $\ket{\psi}$ is a state in Hilbert space $\mathcal{H}=\mathcal{H}_1\otimes..\otimes\mathcal{H}_b$. $\cF_{N,b}^{\rm opt}/\cF_N^{\rm th}$ would then be a reasonable figure of merit for our thermometry protocol.
% The reciprocal of the Fisher information sets a lower bound on the variance of any unbiased estimator of temperature
% \begin{align}
% 	\Delta T^{2}\geq \frac{1}{\QFI{}(T,\rho)}\,,
% \end{align}
% thereby allowing us to quantify the precision of this thermometry scheme. This is known as the Cramer-Rao bound. When the Fisher information is computed using the Gibbs state, we will refer to it as the thermal Cramer-Rao bound.

\section{Indirect measurement through ZZ-interaction}\label{sec:ZZ}
We now consider a simple toy model where the system and ancillas are resonant qubits with bare Hamiltonians $H_{S,A_n} = \hbar\Omega\sigma_{S,A_n}^z$. An interesting first choice of interaction is
\begin{equation}\label{eq:HSA}
    H_{SA_n}^{ZZ} = \frac{\hbar g}{2} \sigma_S^Z\sigma_{A_n}^Z\,.
\end{equation}
This interaction describes the indirect measurement of the system state in the $\ket{g},\ket{e}$ basis \cite{Pellegrini2009,von2018,Seah2019}. In each interaction, the incoming ancilla gets rotated clockwise or counter-clockwise along the $z-$axis at frequency $g/2\pi $ depending on whether the system is in $\ket{g}$ or $\ket{e}$. Since $[U_{SA_n},H_S]=0$ in this case, the fixed point of the map $\Phi$ in Eq.\eqref{eq:rhoSn} remains a Gibbs state ($\rho_S^*=\rho_S^{\rm th}$). This means that one can immediately extract information about the temperature without having to wait for steady state operation.

A natural choice of initial ancilla state would be any state perpendicular to the $Z-$axis, e.g.~$\ket{+_x} = (\ket{g}+\ket{e})/\sqrt{2}$ for maximum distinguishability between the co- and counter-rotating states after interaction. In this case, the Fisher information is dependent on the SA-interaction parameter $g\tau_{SA}$,
\begin{equation}
   \QFI{1}^{\ket{+_{x}}} = \frac{1-\cos(2 g\tau_{SA})}{2}\TFI.
\end{equation}
In the limit of strong $SA$ interaction such that $g\tau_{SA}=\pi/2$, Eq.~\eqref{eq:HSA} describes an ideal pointer measurement where $\ket{+_x}$ becomes a thermal mixture of $\ket{\pm_y}$ (which are orthogonal). The optimal POVM is therefore just a projective measurement in the $y-$basis. In this case, the optimal Fisher information from measuring 1 ancilla is simply the thermal Fisher information $\QFI{1,1}^{\rm opt} = \TFI$. Hereafter, we shall consider the optimal scenario where $g\tau_{SA}=\pi/2$.

With this interaction, the role of the environment is to periodically induce incoherent energy jumps (or ``reset'' the system), such that each incoming ancilla sees a fresh thermal qubit. In the limit $\gamma\tau_{SE}\rightarrow \infty$, one then expects $\cF_N^{\ket{+_x}} \rightarrow N\TFI$.  In the other limit where  $\gamma\tau_{SE}\rightarrow 0$, the system essentially evolves unitarily. If the system is initially in a thermal state, the state measured by each ancilla is the same, i.e. we get a maximally correlated state $\rho_{SA_1...A_N} = p_g \ketbra{g,+_y,...,+_y}{g,+_y,...,+_y}+ p_e \ketbra{e,-_y,...,-_y}{e,-_y,...,-_y}$, where $p_{g,e}$ are the ground and excitation probabilities with $p_g=(\bar{n}+1)/(2\bar{n}+1)$. In this case, the collective measurements of $N>1$ ancillas does not add any information and $\cF_N^{\ket{+_{x}}} \rightarrow \TFI$. 

Hence, the incoming ancillas probe the temperature not only through the thermal probabilities but also through the transitions induced by the environment $p_{x\rightarrow y}$ to bring the system from $\ket{x}$ to $\ket{y}$. Specifically, $p_{g\rightarrow g}=1-(1-e^{-\Gamma})p_e$ and $p_{e\rightarrow g}=(1-e^{-\Gamma}) p_g$ where $\Gamma = \gamma(2\bar{n}+1)\tau_{SE}$ is the effective thermalization rate. We show in Appendix \ref{app:proof} that the quantum Fisher information from measuring $N$ ancillas can be expressed as an arithmetic progression
\begin{eqnarray}\label{eq:QFIN_ZZ}
    \QFI{N}^{\ket{+}}&=&\QFI{1}^{\ket{+}}+(N-1)\Delta,\nonumber\\
    \Delta &=& \left(\frac{\partial \bar{n}}{\partial T}\right)^{2}\sum_{k=g,e}\frac{p_k}{p_{k\rightarrow g}p_{k\rightarrow e}}\left(\frac{\partial p_{k\rightarrow g}}{\partial \bar{n}}\right)^{2}.\nonumber\\
\end{eqnarray}
Here, $\Delta$ captures the additional information gained by subsequent ancillas through the transition probabilities $p_{x\rightarrow y}$. Clearly, for collective measurements on $N\rightarrow\infty$ ancillas,  $\QFI{N}^{\ket{+_{x}}}/N\rightarrow \Delta$. Thus, so long as $\Delta/\TFI>1$, this protocol will beat the thermal Cram\'er-Rao bound. 

\begin{figure}
\centering
\includegraphics[width=\columnwidth]{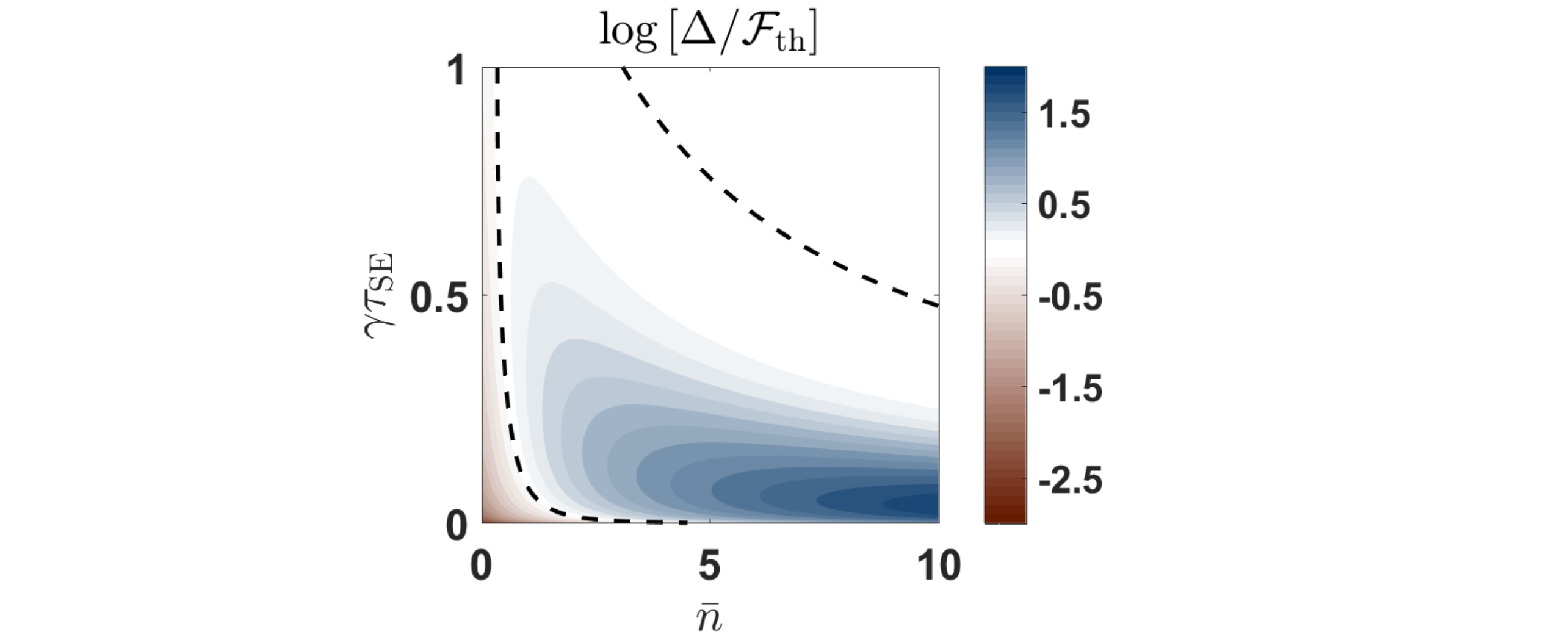}
\caption{\label{fig:Delta} Contour plot of $\Delta/\TFI$ for different $\gamma\tau_{SE}$ and $\bar{n}$. The region within the two black dashed line marks $\Delta>\TFI$.}
\end{figure}

Figure \ref{fig:Delta} shows $\Delta/\TFI$ for different $\gamma\tau_{SE}$ and $\bar{n}$. Here, we see that $\Delta/\TFI$ can be greater than unity especially in the moderately high temperature limit ($\bar{n}\gtrsim 1$). In this limit, since the excitation (and ground) probabilities saturate to $1/2$ with $\TFI\rightarrow 0$, the incoming ancillas can only probe the temperature through the temperature-dependent transition probabilities, captured by $\Delta$. 

\section{Exchange interaction}\label{sec:exchangeint}
In the previous section, we considered a non-invasive indirect measurement of the system's temperature. As a result, the information gained by each ancilla is not independent. As a second choice of interaction, we consider a resonant energy exchange between the system and the incoming ancillas, governed by 
\begin{equation}
    H_{SA_n}^{\rm exc} =\hbar g (\sigma_S^+\sigma_{A_n}^-+ \rm{h.c.}).\label{eq:HExc}.
\end{equation}
In this case, depending on the initial ancilla state, the interaction could serve as an additional incoherent exchange channel that drives $\rho_S^*$ away from a Gibbs state $\rho_S^{\rm th}$ \cite{Strasberg2018,Seah2019,Seah2020}. In particular, if $g\tau_{SA_n}= \pi/2$, $U_{SA}$ describes a full swap where the system is reset to the state of the ancilla, while the ancilla carries away the information. For such a full swap, every SA interaction is like an independent measurement that transfers all the information contained in $\rho_S^*$ to the incoming ancillas. In this case, we are guaranteed at least $\cF_N= N\cF_1$, and we can expect an improvement if the outgoing ancillas are correlated.

In Ref.~\cite{Seah2020}, it was shown that such an interaction allows one to obtain $\QFI{1}^{\ket{g}}>\TFI$ at strong $SA$-couplings where $g\tau_{SA}\approx\pi/2$. Collective enhancement $\QFI{N}^{\ket{\psi}}>N \QFI{1}^{\ket{\psi}}$ was also reported for some single-particle states $\ket{\psi}$, but this happened only at weak $SA$-couplings ($g\tau_{SA}\ll 1$), and the thermal Cram\'er-Rao bound $N \TFI$ was not surpassed in this case. Here, we once again consider $g\tau_{SA}=\pi/2$ and explore strategies that outperform the thermal Cram\'er-Rao bound.

\subsection{Optimal ancilla state for b=1, N=1,2}

\begin{figure}
\centering
\includegraphics[width=\columnwidth]{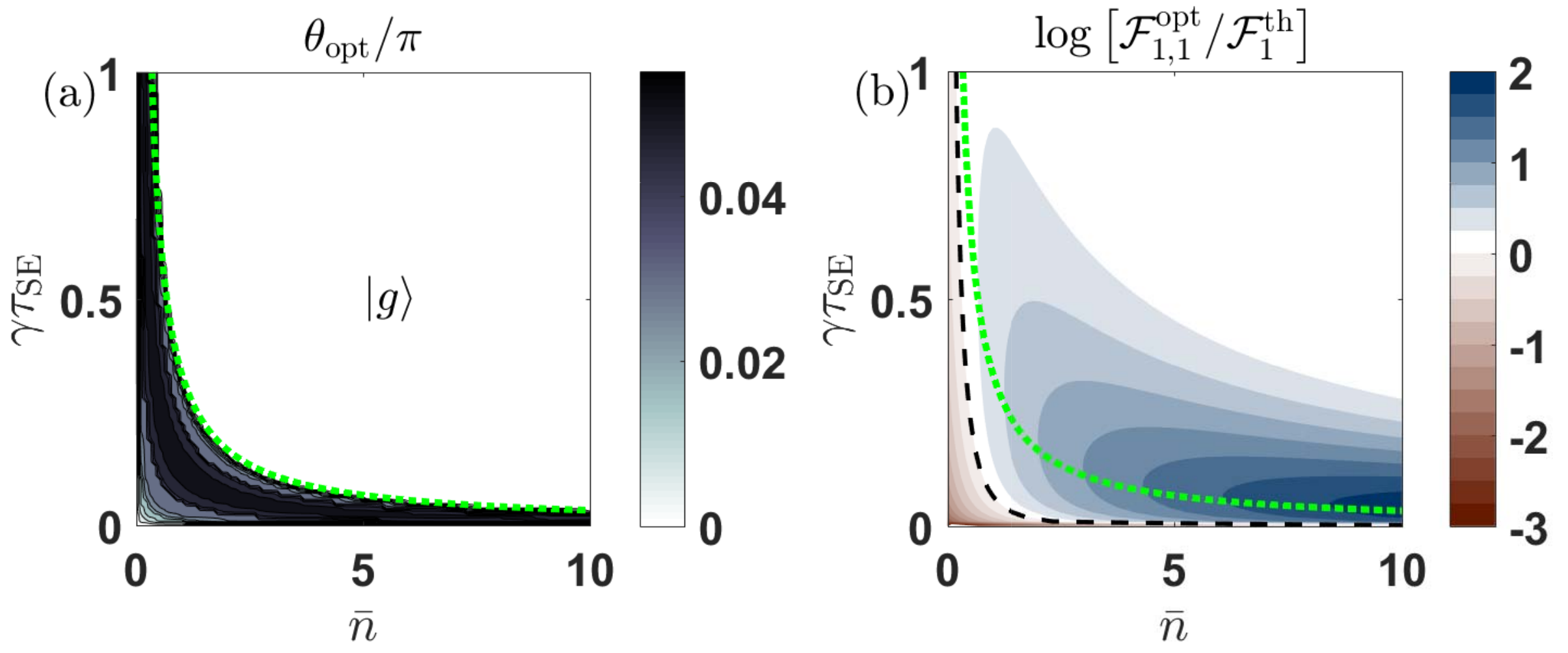}
\caption{\label{fig:b1N1}
Contour plots of (a) optimum angle $\theta$ and (b) corresponding quantum Fisher information for a single ancilla $\QFI{1,1}^{\rm opt}$ normalized to $\TFI$ for different $\bar{n}$ and $\gamma\tau_{SE}$. The green dotted line corresponds to $3\bar{n}\gamma\tau_{SE} = 1$ and the black dashed line in (b) marks $\QFI{1,1}^{\rm opt} = \TFI$.}
\end{figure}

We start by studying the optimisation of $\QFI{1}^{\ket{\psi}}$ over the single-particle input state $\ket{\psi}$ of the ancilla. The exchange interaction $H_{SA}^{\rm exc}$ is invariant under any rotation about the $Z$ axis, thus it is sufficient to consider ancillas whose Bloch vector lies in the $XZ$-plane, i.e.~the single parameter family $\ket{\psi(\theta)}:=\cos(\theta/2)\ket{g}+\sin(\theta/2)\ket{e}$.

Figure \ref{fig:b1N1}(a) shows the optimal state $\ket{\psi(\theta_{\rm opt})}$ for different $\bar{n}$ and $\gamma\tau_{SE}$. Unlike $H_{SA}^{ZZ}$ where the optimal state is the same regardless of temperature, we see here that the optimal state depends on temperature through the $\bar{n}$-dependence. Even though we are unable to predict the exact optimal state (considering temperature is the parameter we are estimating), we see that $\theta_{\rm opt}\ll 1$, i.e.~the optimal state is approximately $\ket{g}$. In fact, for $3\bar{n}\gamma\tau_{\rm SE} \gtrsim 1$ (above green dotted line), the optimal state is always $\ket{g}$. Elsewhere, while the optimal state deviates slightly from $\ket{g}$, we have verified that $\QFI{1}^{\ket{g}}\geq 0.9957 \QFI{1,1}^{\rm opt}$. In addition, it is advantageous to use $\ket{g}$ as ancillas since the state $\rho_{SA_1...A_N}$ is always diagonal, implying that the optimal POVM would just be projective measurements in the $Z-$basis regardless of the number of ancillas measured.

%\replaced[id=a]{As temperature is the very parameter this thermometry protocol aims to estimate, it is not possible to prepare the ancilla in the state $\ket{\psi(\theta_{\rm opt})}$ before running the protocol}
% {Since the temperature is the parameter we are measuring, we would generally not know $\ket{\psi(\theta_{\rm opt})}$ beforehand.}
% However, this is not a major issue \deleted[id=a]{here }as we see that 

Here, we also see that compared to the standard measurement interaction $H_{SA}^{ZZ}$, an exchange interaction like $H_{SA}^{\rm exc}$ actually allows us to beat the thermal Cram\'er-Rao bound with just one ancilla, especially in the high-temperature limit. As an example, when $\bar{n}=10$, the maximum quantum Fisher information from just one ancilla is $\QFI{1,1}^{\rm opt} \approx 77.3\TFI$. Correspondingly, for $H_{SA}^{ZZ}$, the maximum $\Delta$ is approximately $71.8\TFI$, attainable only for collective measurements $N\gg1$.
%$\QFI{1}^{\ket{\psi(\theta)}}=\QFI{\rm opt}(1,1)$, as a function of $\bar{n}$ and $\gamma\tau_{SE}$, where $\QFI{\rm opt}(b,N)$ stands for the optimum quantum Fisher information achievable for ancilla of block size $b$ with $N$ of them collectively measured. 

% At first, this graph may look unpleasant: the optimal state depends on $\bar{n}$ (i.e. on $T$, the parameter to be measured) which cannot be known during the preparation of the ancilla. A closer look reveals however that the ground state $\ket{g}$ ($\theta=0$), which was used in the case study of Ref.~\cite{Seah2020}), is a relatively robust choice. Indeed, that state is always optimal in the region $3\bar{n}\gamma\tau_{SE}\gtrsim 1$. Elsewhere, while the optimal state is different, we have verified that $\QFI{1}^{\ket{g}}\geq 0.9957 \QFI{1}^{\rm opt}$.

In Fig.~\ref{fig:b1N1}(b), we see that the thermal Cram\'er-Rao bound can be beaten for almost all temperatures by the use of $\ket{g}$, so long as we choose a suitable time of interaction with the environment $\tau_{SE}$. These choices may be critical for $\bar{n}\ll 1$; as soon as $\bar{n}\gtrsim 1$, $\QFI{1}^{\ket{g}}>\TFI$ holds for almost any value of $\tau_{SE}$ as well. The best advantage is found for large $\bar{n}$ (high temperature) and low $\gamma\tau_{SE}$. This is not entirely surprising as $\rho^*_S\approx\rho^{\rm{th}}$ for large $\gamma\tau_{SE}$, rendering the scheme ineffective in this parameter regime. To gain more insight on the enhancement, notice that the use of ground state ancillas at full swap brings the steady state of the system to 
\begin{equation} \label{eq:systemsteady}
    \rho_S^*=(1-\frac{\bar{n}(1-e^{-\Gamma})}{2\bar{n}+1})\ket{g}\bra{g}+\frac{\bar{n}(1-e^{-\Gamma})}{2\bar{n}+1}\ket{e}\bra{e}.
\end{equation}
Similar to the $H_{SA}^{ZZ}$, the ancillas are able to probe the environment not only through the excitation probability but also via the effective thermalization rate $\Gamma$, thus contributing to a higher Fisher information. However, unlike $H_{SA}^{ZZ}$ which does not perturb $\rho_S^*$ away from $\rho_S^{\rm th}$, a full swap interaction ensures that any information about temperature contained in $\rho_S^*$ is carried away by each incoming ancilla and for $\ket{g}$ which does not have coherences to build up additional correlations, $\mathcal{F}^{\ket{g}}_{N}=N\mathcal{F}^{\ket{g}}_{1}$.

\begin{figure}
\centering
\includegraphics[width=\columnwidth]{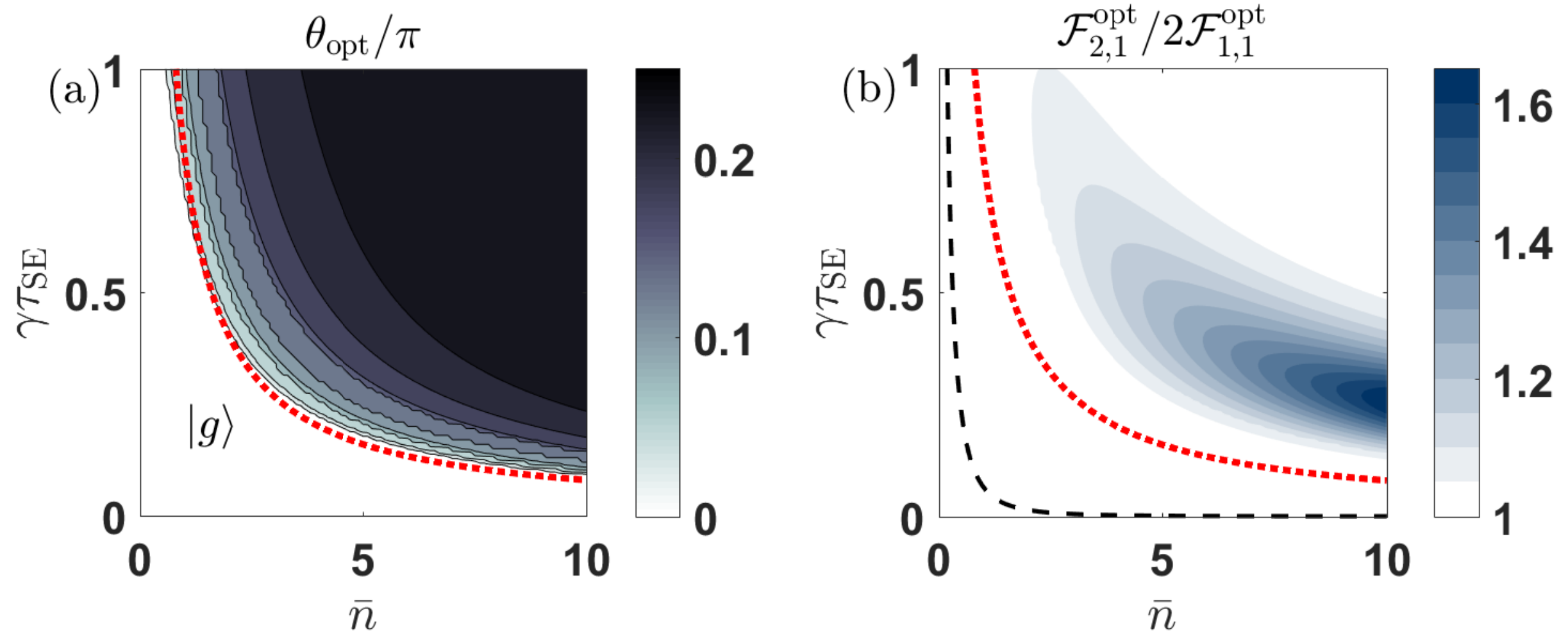}
\caption{\label{fig:b1N2}
Contour plots of (a) optimum angle $\theta$ and (b) corresponding quantum Fisher information for two single-particle ancilla $\QFI{2,1}^{\rm opt}$ normalized to that of one single-particle ancilla $\QFI{1,1}^{\rm opt}$ for different $\bar{n}$ and $\gamma\tau_{SE}$. The red dotted line $\bar{n}\gamma\tau_{SE} = 0.8$ marks the approximate region where the optimum state is $\ket{g}$ and the black dashed line in (b) marks $\QFI{2,1}^{\rm opt}= \TFI$.}
\end{figure}

Once we allow for collective measurements on multiple ancillas, we find that the ground state $\ket{g}$ is no longer optimum. Specifically, for $N=2$, Fig.~\ref{fig:b1N2}(a) shows that the optimal angle $\theta_{\rm opt}$ can now take values up to $\pi/4$ in the limit $\bar{n}\gamma\tau_{SE}\gg 1$.
While the excitation probability decays to $\bar{n}/(2\bar{n}+1)$ at a rate $\Gamma$, the coherences $\ketbra{g,g}{g,e}$ and $\ketbra{g,g}{e,g}$ scale like $e^{-\Gamma/2}\sin{2\theta}$, i.e.~apart from near-zero information gained through the ground and excited probabilities independent of $\theta$, one also gains additional information about $\Gamma$ through the vanishing coherences, which peaks at $\theta=\pi/4$ and decays at half the rate.

Even though there is a collective advantage when $\bar\gamma\tau_{SE}\approx0.8$ (dotted red), one can always choose a smaller $\gamma\tau_{\rm SE}$ for a given $\bar{n}$ such that the $\ket{g}$ still yields a higher Fisher information. As an example, the maximum enhancement in Fig.~\ref{fig:b1N2}(b) occurs when $\gamma\tau_{SE} \approx 0.26$ and $\bar{n} = 10$ where $\QFI{2,1}^{\rm opt}/2\QFI{1,1}^{\rm opt} \approx 1.65$ and $\QFI{2,1}^{\rm opt}/2\TFI \approx 3.6$. However, in Fig.~\ref{fig:b1N1}(b), we see that for the same $\bar{n}$, one can already attain $\QFI{1}^{\ket{g}}/\TFI \approx 100 $ even at a considerably smaller $\gamma\tau_{\rm SE} \approx 0.04$. In other words, ground state ancillas alone are sufficient for probing the environment even in the limit of weak system-environment coupling where $\gamma\tau_{\rm SE}\ll 1$. 

However, this advantage from collective measurements decreases with increasing $N$, with $\QFI{3,1}^{\rm opt}/3\QFI{1,1}^{\rm opt} \leq 1.93$ and $\QFI{4,1}^{\rm opt}/4\QFI{1,1}^{\rm opt} \leq 2.06$. Therefore, it suffices to consider say $N=1,2$ at $g\tau_{SA}=\pi/2$ so long as we are not restricted by the interaction strength or duration, especially since collective measurements for large $N$ described by the optimal POVM $\Pi_x$ may not always be feasible to implement.

\subsection{Optimal ancilla state for b=2, N=2,4}

\begin{figure}
\centering
\includegraphics[width=\columnwidth]{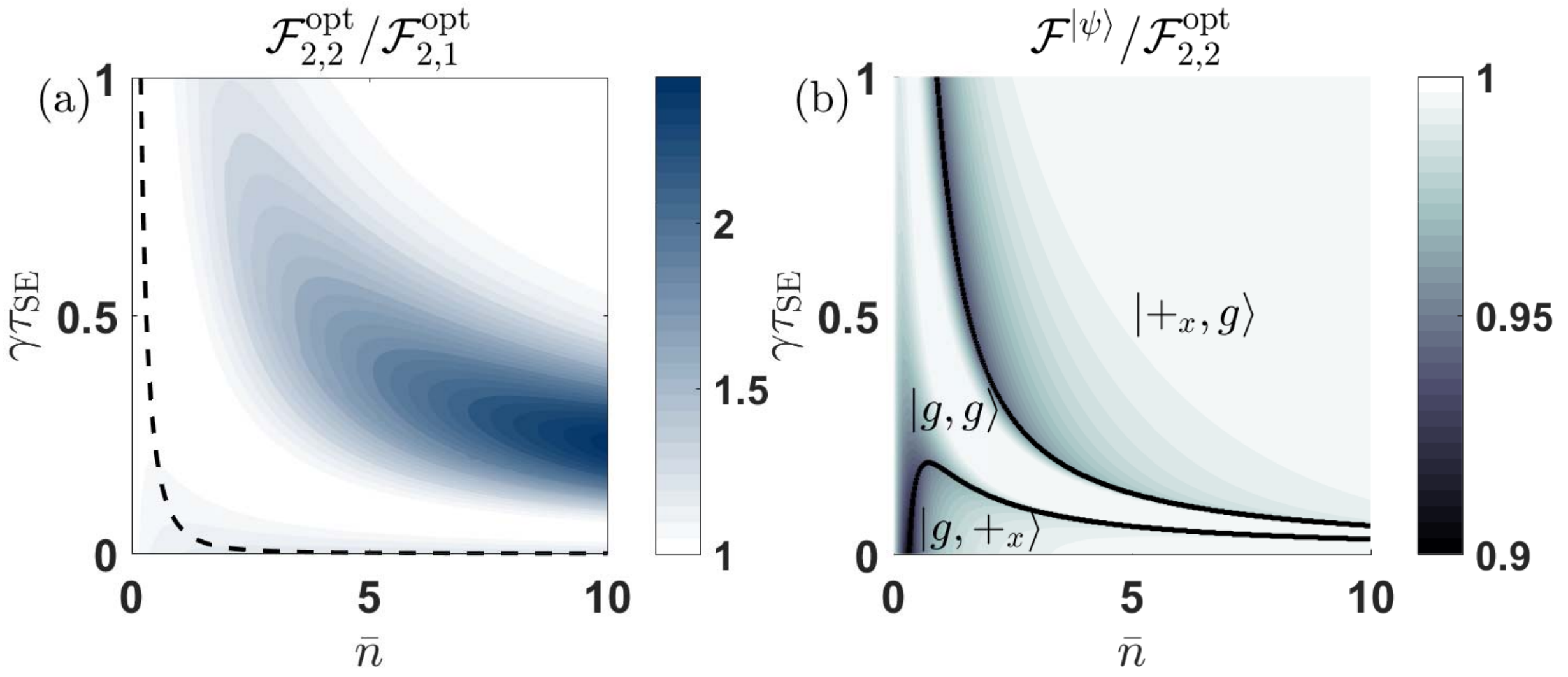}
\caption{\label{fig:b2N2} (a) Contour plot of ratio $\QFI{2,2}^{\rm opt}/\QFI{2,1}^{\rm opt}$, the optimal Fisher information for $b=2$, $N=2$ normalized to that of $b=1$, $N=2$, for a given $\gamma\tau_{SE}$ and $\bar{n}$ and (b) the ratio $\QFI{2}^{\ket{\psi}}/\QFI{2,2}^{\rm opt}$,  the Fisher information of \emph{almost-}optimal states $\ket\psi = \ket{+_x,g},\ket{g,g},\ket{g,+_x}$ normalized to the optimal Fisher information for $b=2$, $N=2$. The black dashed line in (a) marks both $\QFI{2,1}^{\rm opt}= 2\TFI= 2\TFI$.}
\end{figure}

We parametrise the family of two-particle ancilla states by their Schmidt decomposition
\begin{equation}\label{eq:2qubitstate}
    \ket{\psi(r,\vec{m},\vec{n},\alpha)}:=\sqrt{r}\ket{+_{m},+_{n}}+e^{i\alpha}\sqrt{1-r}\ket{-_{m},-_{n}}
\end{equation} where for $k =m,n$ we write
\begin{eqnarray}\label{eq:blochstate}
     \ket{+_k} &=& \cos \frac{\theta_k}{2} \ket{g} + e^{i\phi_k}\sin\frac{\theta_k}{2} \ket{e} \nonumber\\
     \ket{-_k} &=& e^{-i\phi_k}\sin \frac{\theta_k}{2} \ket{g} - \cos\frac{\theta_k}{2}\ket{e}.
\end{eqnarray}
Again, due to the invariance of $H_{SA}^{\rm exc}$ under rotations about $Z$, it is sufficient to consider the five parameters $(r,\theta_m,\theta_n,\phi_n,\alpha)$ while setting $\phi_m = 0$.

Figure~\ref{fig:b2N2}(a) compares the optimal quantum Fisher information $\QFI{2,2}^{\rm opt}/\QFI{2,1}^{\rm opt}$ from collective measurement on two ancillas for block sizes $b=1,2$. We see that $\QFI{2,2}^{\rm{opt}}$ can exceed twice  $\QFI{2,1}^{\rm{opt}}$, allowing us to further surpass the thermal Cram\'er-Rao bound. Here, the states that give $\QFI{2,2}^{\rm opt}$ are nearly uncorrelated, with $r\geq 0.999994$, thereby indicating within numerical precision that, at strong interaction, initial entanglement between ancillas is not required for metrological advantage.

Similarly to the case of $b=N=1$, we could identify simple product states that are close to optimal -- only here, instead of just one state, we need three: $\ket{g,g},\ket{+_{x},g}$ and $\ket{g,+_x}$. Before presenting the evidence for this, let us highlight that $\ket{+_{x},g}$ and $\ket{g,+_{x}}$ describe the same stream of ancillas, one alternating $\ket{g}$ and $\ket{+_x}$; they differ only by when the two-body collective measurement is performed.

%because the collective measurement is done $\ket{+_{x},g}$ have the collective measurement done after $\ket{g}$ interacts with the system whereas for $\ket{g,+_{x}}$, it occurs after $\ket{+_{x}}$

The parameter regions in which each of these states is almost optimal is plotted in Fig.~\ref{fig:b2N2}(b). Specifically, for each region marked by states $\ket{\psi}=\ket{g,g}$, $\ket{g,+_x}$ and  $\ket{+_{x},g}$, $\QFI{2}^{\ket{\psi}}/\cF_{2,2}^{\rm opt} \geq 0.909$, $0.921$ and $0.931$ respectively. Here, $\rho_S^*$ for both $\ket{+_x,g}$ and $\ket{g,g}$ is given by \eqref{eq:systemsteady}. This implies that information gained by the first ancilla via a full swap would be the same for both $\ket{+_x}$ or $\ket{g}$, since they interact with the same system state. However, should the system be perturbed to $\ket{+_x}$, the second ancilla will gain additional information from the decaying coherences, especially in the high-temperature limit where the ground and excited probabilities saturate to $1/2$. 

From the results so far, we see that the thermometry protocol following $H_{SA}^{\rm exc}$ gives us significant advantage in the moderate-to-high temperature limit ($\bar{n}\gtrsim 1$), but does not beat the Cram\'er-Rao bound in the low-temperature limit ($\bar{n}\ll 1$). Considering $b=N=2$ for instance, one needs to go to strong and long SE interactions ($\gamma\tau_{SE}\geq 1$) in order to beat the Cram\'er-Rao bound at $\bar{n}\leq 0.189$. Furthermore, we also verified numerically that there is no significant enhancement even if we consider collective measurements on $N>2$ ancillas.

\section{Conclusion}
In this paper, we have explored the ways that a collisional thermometry protocol can overcome the thermal Cram\'er-Rao bound. Both the $ZZ-$interaction and the exchange interaction return significant enhancements in the moderate to high temperature limit ($\bar{n}\gtrsim1$). The $ZZ-$interaction is beneficial in that the strategy that returns the optimal Fisher information is independent of the temperature of the system being measured. However, one can only surpass the thermal Cram\'er-Rao bound when measuring at least two ancillas collectively. The exchange interaction, on the other hand, does not require collective enhancement for significant enhancements over $\TFI$. Furthermore, an advantage is observed over a larger parameter region with the energy exchange channel. Both interactions display collective advantage although this advantage diminishes as the number of ancilla increases. For blocks of two ancillas, we further verified that entangled states do not outperform some product ones.

\acknowledgments{
\emph{Acknowledgments --} We thank Stefan Nimmrichter for discussions in the early phase of this project. This research is supported by the National Research Foundation and the Ministry of Education, Singapore, under the Research Centres of Excellence programme.}

\appendix
\begin{widetext}

\section{Collective enhancement for ZZ interaction}\label{app:proof}

\begin{figure}
    \centering
    \includegraphics[width=0.6\columnwidth]{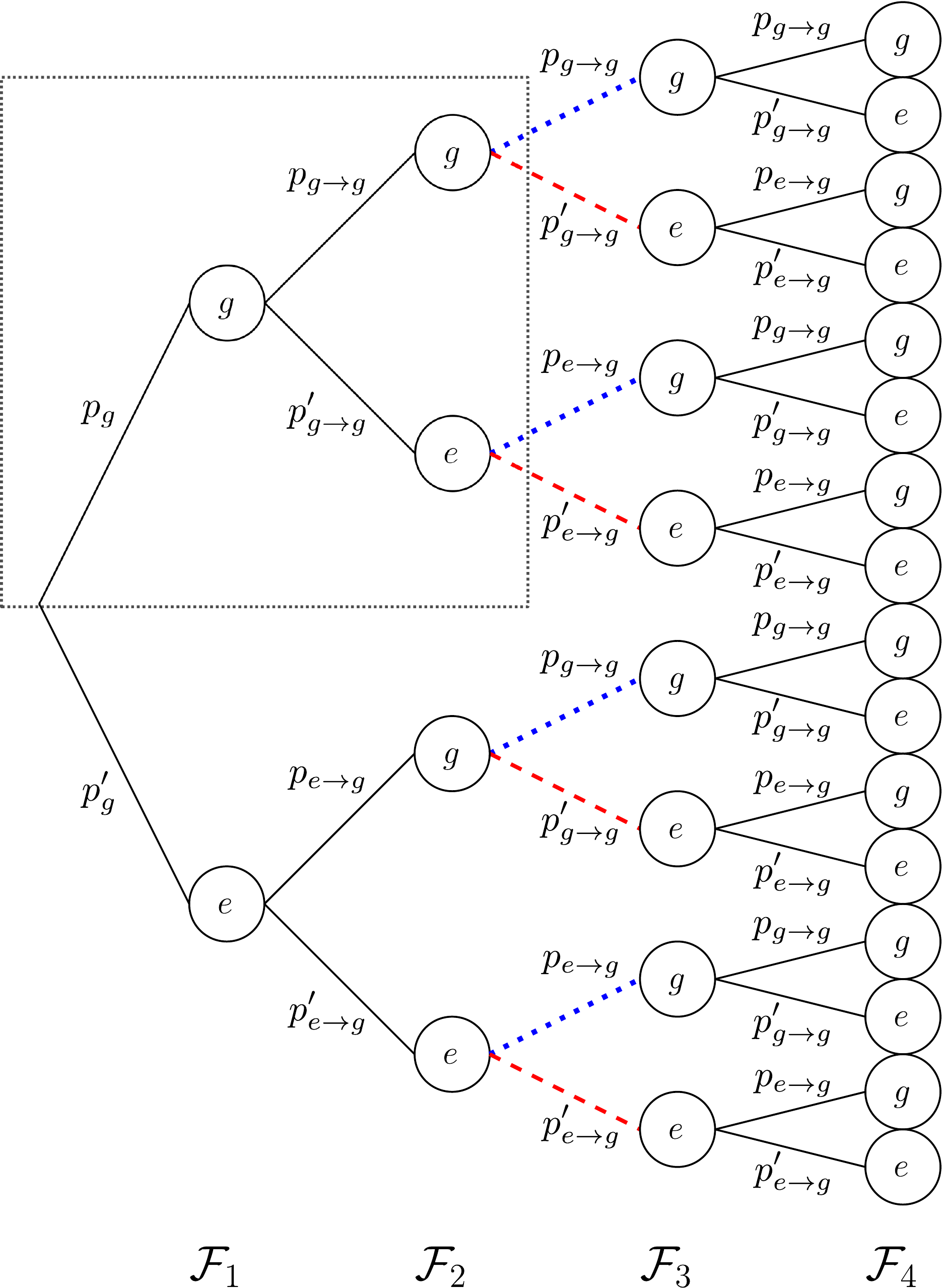}
    \caption{Probability tree diagram of the measurement outcome of 4 ancilla.}
    \label{fig:TreeDiag}
\end{figure}

In this appendix, we prove that the collective advantage for $H_{SA}^{ZZ}$ takes the form of a linear progression
\begin{eqnarray}\label{eq:QFIzz}
    \QFI{N}^{\ket{+}}&=&\QFI{1}^{\ket{+}}+(N-1)\left(\frac{\partial \bar{n}}{\partial T}\right)^{2}\sum_{k=g,e}\frac{p_k}{p_{k\rightarrow g}p_{k\rightarrow e}}\left(\frac{\partial p_{k\rightarrow g}}{\partial \bar{n}}\right)^{2}.
\end{eqnarray}

To this end, we consider a probability tree in Fig.~\ref{fig:TreeDiag} the measurement outcomes with 4 ancillas. Here, the ground state probability is $p_g$ and $p_{g\rightarrow g}$ and $p_{e\rightarrow g}$ are the transition probabilities of staying in the ground state and jumping from excited state to ground state respectively, and $p'_{i}=1-p_{i}$.

To reduce notational clutter, we define two functions: $f(x)=x\brac{\diff{\ln x}}^{2}=\frac{1}{x}\brac{\diff{x}}^{2}$ and $g(x,y)=f(xy)+f(xy')$. The first function allows us to reexpress QFI of one ancilla as $\QFI{1}=f(p_g)+f(p_g')$, and the second function simplifies the expression of the elements of the Fisher information in the red box in Fig.~\ref{fig:TreeDiag} down to $g(p,p_{g\rightarrow g})$. One can easily show that 
\begin{equation}
    	g(p_g,p_{g\rightarrow g})=f(p_g)+\frac{p_g}{p_{g\rightarrow g}p_{g\rightarrow g}'}\brac{\diff{p_{g\rightarrow g}}}^{2},
\end{equation}
and therefore
\begin{eqnarray}
\QFI{2}&=&f(p_gp_{g\rightarrow g})+f(p_gp'_{g\rightarrow g})+f(p'p_{e\rightarrow g})+f(p_g'p'_{e\rightarrow g})\nonumber\\
&=&g(p_g,p_{g\rightarrow g})+g(p_g',p_{e\rightarrow g})\nonumber\\
&=&\underbrace{f(p_g)+f(p_g')}_{\QFI{1}}+\frac{p_g}{p_{g\rightarrow g}p_{g\rightarrow g}'}\brac{\diff{p_{g\rightarrow g}}}^{2}+\frac{p_g'}{p_{e\rightarrow g}p_{e\rightarrow g}'}\brac{\diff{p_{e\rightarrow g}}}^{2}.
\end{eqnarray}

With this notation, $\QFI{3}$ can be calculated as
\begin{eqnarray}
\QFI{3}&=&g(p_gp_{g\rightarrow g},p_{g\rightarrow g})+g(p_gp'_{g\rightarrow g},p_{e\rightarrow g})+g(p'p_{e\rightarrow g},p_{g\rightarrow g})+g(p_g'p'_{e\rightarrow g},p_{e\rightarrow g})\nonumber\\
&=&\underbrace{f(p_gp_{g\rightarrow g})+f(p_gp'_{g\rightarrow g})+f(p_g'p_{e\rightarrow g})+f(p_g'p'_{e\rightarrow g})}_{\QFI{2}}\nonumber\\&+&\frac{p_gp_{g\rightarrow g}+p_g'p_{e\rightarrow g}}{p_{g\rightarrow g}p'_{g\rightarrow g}}\brac{\diff{p_{g\rightarrow g}}}^{2}+\frac{pp'_{g\rightarrow g}+p_g'p'_{e\rightarrow g}}{p_{e\rightarrow g}p'_{e\rightarrow g}}\brac{\diff{p_{e\rightarrow g}}}^{2}\nonumber\\
&=&\QFI{2}+\frac{p}{p_{g\rightarrow g}p_{g\rightarrow g}'}\brac{\diff{p_{g\rightarrow g}}}^{2}+\frac{p'}{p_{e\rightarrow g}p_{e\rightarrow g}'}\brac{\diff{p_{e\rightarrow g}}}^{2},
\end{eqnarray}

where the last equality comes from substituting the expression for  $p_{g\rightarrow g}=\Lambda p'+1$ and $p_{e\rightarrow g}=\Lambda p_{g\rightarrow g}$.

Following the same methodology, we obtain
\begin{eqnarray}\label{eq:appF4}
	\QFI{4}&=&\QFI{3}+\frac{\color{blue}p_gp_{g\rightarrow g}p_{g\rightarrow g}+p_gp_{g\rightarrow g}'p_{e\rightarrow g}+p_g'p_{g\rightarrow g}p_{g\rightarrow g}+p_g'p'_{g\rightarrow g}p_{e\rightarrow g}}{p_{g\rightarrow g}p'_{g\rightarrow g}}\brac{\diff{p_{g\rightarrow g}}}^{2}\nonumber\\
	&+&\frac{\color{red}p_gp_{g\rightarrow g}p'_{g\rightarrow g}+p_gp'_{g\rightarrow g}p'_{e\rightarrow g}+p_g'p_{g\rightarrow g}p'_{g\rightarrow g}+p_g'p'_{g\rightarrow g}p'_{e\rightarrow g}}{p_{e\rightarrow g}p'_{e\rightarrow g}}\brac{\diff{p_{e\rightarrow g}}}^{2}\nonumber\\
	&=&\QFI{3}+\frac{p}{p_{g\rightarrow g}p_{g\rightarrow g}'}\brac{\diff{p_{g\rightarrow g}}}^{2}+\frac{p'}{p_{e\rightarrow g}p_{e\rightarrow g}'}\brac{\diff{p_{e\rightarrow g}}}^{2}.
\end{eqnarray}

The probability terms in blue (second term of first line of \eqref{eq:appF4}) is the probability that the third ancilla will be found in the ground state, represented on Fig.~\ref{fig:TreeDiag} by the blue dotted line.  This is just $p_g$ as the protocol is merely an ideal pointer measurement of the system, which is always in the Gibbs state. Similarly, the terms in red (term on second line of \eqref{eq:appF4}) come from the dashed dotted lines in Fig.~\ref{fig:TreeDiag} and will reduce to $p_g'$. Therefore, we can conclude by induction that Eq.~\eqref{eq:QFIzz} is true.
\end{widetext}
\end{document}